\documentclass[conference]{IEEEtran}
\IEEEoverridecommandlockouts
\usepackage{amsmath,amsfonts}
\usepackage[ruled,vlined]{algorithm2e}
\usepackage{tabularx}
\usepackage{subfig}
\usepackage{graphicx}
\usepackage{physics}
\usepackage{textcomp}
\usepackage{xcolor}
\usepackage{fancyhdr}
\usepackage{enumitem}
\usepackage{afterpage}
\usepackage{array}
\usepackage{bm}
\usepackage{float}
\usepackage[normalem]{ulem}
\usepackage[numbers,sort&compress]{natbib}
\usepackage{caption}
\usepackage{environ}
\usepackage{textcomp}
\usepackage{booktabs}
\usepackage{makecell}
\usepackage{url}
\usepackage{tcolorbox,tikz}
\tcbuselibrary{skins}
\tcbuselibrary{breakable}
\newtcolorbox{mybox}[3][]
{
  breakable, 
  enhanced,
  colback         = #2!10,
  colframe        = #2!5,
  boxsep          =-0.5mm,
  borderline west = {1.0mm}{0.05mm}{#3!30}, 
  borderline north = {0.3mm}{0.05mm}{#3!30}, 
  borderline east = {0.3mm}{0.05mm}{#3!30}, 
  borderline south = {0.3mm}{0.05mm}{#3!30}, 
  #1,
}

\newcommand{\sol}{\textsc{Sprout}}
\newcommand{\oracle}{\textsc{Oracle}}
\newcommand{\base}{\textsc{Base}}
\newcommand{\cooopt}{CO$_2$\_\textsc{Opt}}
\newcommand{\modelopt}{\textsc{Model}\_\textsc{Opt}}
\newcommand{\static}{\sol{}\_\textsc{Sta}}

\usepackage{pifont}% http://ctan.org/pkg/pifont
\newcolumntype{P}[1]{>{\centering\arraybackslash}p{#1}}
\colorlet{shadecolor}{gray!20}
\usepackage{tikz}
\newcommand*\blackcircled[1]{\tikz[baseline=(char.base)]{
            \node[shape=circle,draw,inner sep=0.5pt,fill=black,text=white] (char) {#1};}}

\newtheorem{definition}{Definition}

\def\BibTeX{{\rm B\kern-.05em{\sc i\kern-.025em b}\kern-.08em
    T\kern-.1667em\lower.7ex\hbox{E}\kern-.125emX}}

\begin{document}
\author{
        % Primary authors
        Baolin Li\IEEEauthorrefmark{1},        
        Yankai Jiang\IEEEauthorrefmark{1},
        Vijay Gadepally\IEEEauthorrefmark{2},
        Devesh Tiwari\IEEEauthorrefmark{1}\\
    \IEEEauthorrefmark{1} Northeastern University,
    \IEEEauthorrefmark{2} MIT}

\title{Toward Sustainable GenAI using Generation Directives for Carbon-Friendly Large Language Model Inference}
% \subtitle{Paper Type: Open-source tools or data}


\maketitle

\begin{abstract}

The rapid advancement of Generative Artificial Intelligence (GenAI) across diverse sectors raises significant environmental concerns, notably the carbon emissions from their cloud and high performance computing (HPC) infrastructure. This paper presents \sol{}, an innovative framework designed to address these concerns by reducing the carbon footprint of generative Large Language Model (LLM) inference services. \sol{} leverages the innovative concept of ``generation directives" to guide the autoregressive generation process, thereby enhancing carbon efficiency. Our proposed method meticulously balances the need for ecological sustainability with the demand for high-quality generation outcomes. Employing a directive optimizer for the strategic assignment of generation directives to user prompts and an original offline quality evaluator, \sol{} demonstrates a significant reduction in carbon emissions by over 40\% in real-world evaluations using the Llama2 LLM and global electricity grid data. This research marks a critical step toward aligning AI technology with sustainable practices, highlighting the potential for mitigating environmental impacts in the rapidly expanding domain of generative artificial intelligence.
\end{abstract}

\section{Introduction}
\label{sec:intro}

% AI boom, carbon concern

The emergence of Generative Artificial Intelligence (GenAI) has significantly impacted various sectors such as scientific discovery, engineering, law, and finance~\cite{jumper2021highly,pierce2023law,chen2023fiction}, signaling a major shift in how challenges and tasks are approached in these fields. This technology's ability to produce novel content from existing data has cemented its popularity in datacenters worldwide. However, the AI boom, driven by the demand for GenAI, has prompted concerns over its environmental impact, particularly in terms of carbon emissions associated with the energy-intensive nature of these technologies. OpenAI's reported pursuit of trillions in investment for AI chips~\cite{samaltman}, destined for cloud and high performance computing (HPC) datacenters, underscores the scale of infrastructure expansion required to support GenAI's growth. With global datacenter energy consumption projected to more than double from 460 TWh in 2022 to 1000 TWh by 2026~\cite{iea}, the consequent surge in electricity generation to power these facilities could contribute to 8\% of global carbon emissions within a decade~\cite{gupta2022act}, highlighting the urgent need for sustainable practices in the rapidly expanding realm of artificial intelligence.

Generative Large Language Models (LLMs), such as ChatGPT, experience over 1 billion visits monthly, underscoring an urgent need for research focused on minimizing their environmental impact. Training these models requires extensive compute cycles and corresponding carbon footprint. However, it is the inference processes of these LLMs that are poised to become the predominant source of emissions, according to various prior studies~\cite{chien2023reducing,wu2022sustainable,de2023growing}. Unlike traditional natural language understanding models that predict a single masked word or sentiment, generative LLMs are even more carbon-demanding as they perform iterative predictions for each request until reaching a predefined token or iteration limit. Despite the critical nature of this issue, there's a noticeable gap in research dedicated to reducing carbon emissions specifically from the inference operations of generative language models. Addressing this gap is crucial for making GenAI advancements sustainable and environmentally responsible.

In this paper, we design \sol{} as the first work to address the sustainability challenges in running a generative LLM inference service. Various previous works have attempted to reduce the carbon footprint of machine learning (ML) applications in cloud and HPC datacenters~\cite{wu2022sustainable,li2023clover,anderson2023treehouse,dodge2022measuring}, but none has designed optimizations tailored to LLM inference which is becoming a dominant workload in information technology and requires immediate intervention to reduce its carbon footprint. The following summarized the insights behind \sol{} and its key contributions.  \vspace{2mm}

\noindent\textbf{Introduction of generation directives to LLM inference for carbon saving. } Previous works have identified the opportunity to manipulate the number of parameters in the model to save energy and carbon~\cite{wan2020alert,li2023clover,romero2021infaas}, while \sol{} is the first work to identify that in generative language model inference, its autoregressive generation pattern presents a unique opportunity outside of the dimension that any previous work has explored. \sol{} introduces the concept of ``generation directives'', a strategy to indirectly manipulate the number of autoregressive inference iterations while providing high-quality content generation. For example, a directive can guide the model to provide a concise response, saving significant carbon from generating a long sequence while still being accurate. Identifying the variability in the carbon intensity of the electricity generation and the diverse requirements of different tasks, \sol{} can leverage different generation directives to minimize the carbon footprint of LLM inference with a guarantee of generation quality. \vspace{2mm}

\noindent\textbf{Design and implementation of carbon-optimized generation directive configuration for LLM inference. } We present \sol{}, an innovative carbon-aware generative language model inference framework designed to reduce carbon footprint through the strategic use of token generation directives while maintaining high-quality outputs. From the selection of directive levels based on electricity grid carbon intensity and user behavior variability, \sol{} introduces a linear programming approach for system-level optimization, balancing carbon savings with generation quality. \sol{} identifies the difficulty in retrieving generation quality feedback, and implements an automatic offline quality assessment mechanism to ensure the framework's decisions are informed by up-to-date generation quality. \vspace{2mm}

\noindent\textbf{Evaluation of \sol{} with real-world LLM and electricity grid operators. } Our extensive evaluation of the \sol{} system demonstrates its effectiveness in reducing carbon emissions of LLM inference services by more than 40\% while still achieving high generation quality. We evaluate the system using production software setup, the latest open-source Llama2 LLM, representative corpus to synthesize user prompts, and real carbon intensity traces from multiple global electricity grid operator regions. Our real-system prototype demonstrates that \sol{} is superior to its competitors and is well-aligned to a hypothetical yet unattainable \oracle{} scheme. These results suggest \sol{} offers a meaningful step forward in making LLM inference systems more environmentally friendly, contributing to the ongoing effort to align GenAI technology with sustainable practices.

\section{Background and Motivation}

\subsection{Background}
\label{sec:bkgd}

\noindent\textbf{Carbon footprint of an inference request. } The carbon footprint is a metric for quantifying the amount of greenhouse gas emissions (gCO$_2$) generated. When requesting a service from a cloud and HPC datacenter (e.g., HTTP requests), its carbon footprint comprises the operational carbon and embodied carbon. The operational carbon comes from the electricity grid that powers the datacenter, which powers the hardware (e.g., GPUs) that serves the request. The carbon intensity (denoted as $CO_2^{\text{Intensity}}$) of the grid, representing the amount of CO$_2$ emission per energy usage (gCO$_2$/kWh), reflects the ``greenness'' of the energy source. For example, wind turbines have lower carbon intensity than coal power plants. Due to the difference in availability and stability of renewable energy, carbon intensity varies significantly over time and across geographical regions. The carbon intensity is the multiplier to the request's energy consumption when quantifying its operational carbon footprint.

% Embodied carbon (denoted as $CO_2E$) is the carbon emission incurred during the manufacturing and packaging of computer components, thus it is ``embodied'' into the device. For an inference request that runs on the device, its portion share of the embodied carbon should be its execution time as a portion of the device's lifetime. Please refer to previous works for a more detailed explanation of embodied and operational carbon~\cite{gupta2022act,li2023toward}. 
Embodied carbon (denoted as $CO_2^{\text{Embed}}$) represents the carbon emissions associated with the manufacturing and packaging of computer components, effectively ``embodied" within the device itself. For an inference request processed by a computing device, its share of embodied carbon is proportional to the execution time relative to the device's overall operational lifespan. More detailed information about the embodied and operational carbon footprint in sustainable computing is available in previous works~\cite{gupta2021chasing,gupta2022act}. The total carbon footprint of serving an inference request, $C_{\text{req}}$, can be formally expressed as:
\begin{align}
\label{eq:bkgd}    
C_{\text{req}}=CO_2^{\text{Intensity}}\cdot E_{\text{req}} + \frac{CO_2^{\text{Embed}}}{T_{\text{life}}}\cdot T_{\text{req}}
\end{align}
Here, $E_{\text{req}}$ and $T_{\text{req}}$ represent the energy consumption and execution time for the request, respectively, with $T_{\text{life}}$ indicating the assumed device lifespan, set to five years for this analysis. Given that the lifespan of the device significantly exceeds any single request's execution time, operational carbon predominantly dictates the total carbon footprint, except in scenarios where $CO_2^{\text{Intensity}}$ approaches zero.
\begin{figure}[t]
    \centering
    \includegraphics[scale=0.34]{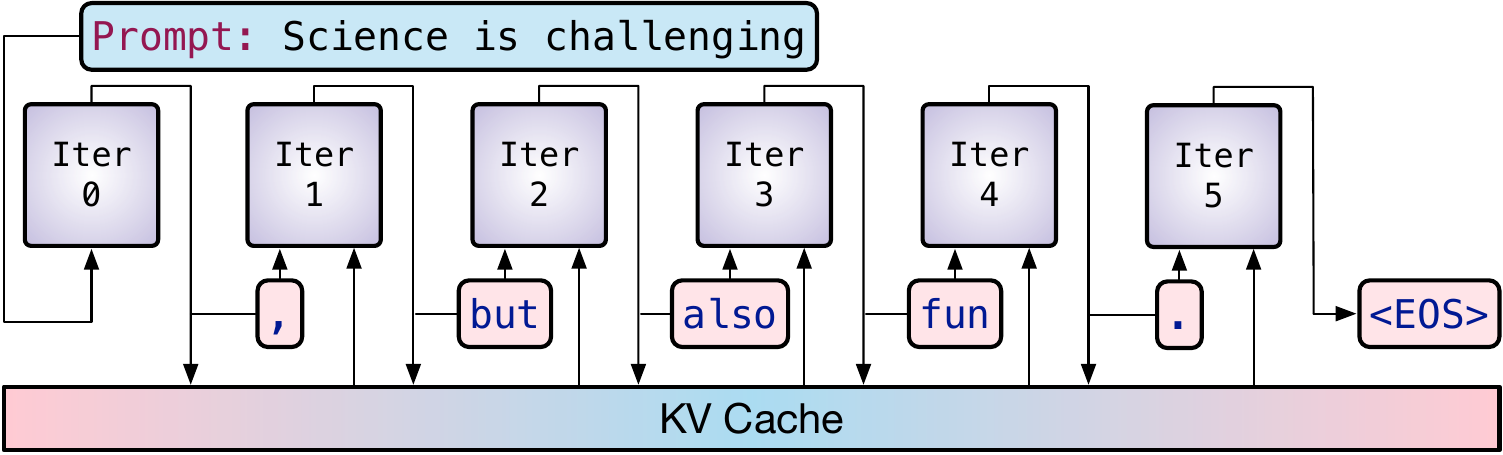}%\newline
    \vspace{1mm}
    \hrule
     \vspace{1mm}
    \caption{The auto-regressive generation process of generative language model inference.}
    % \vspace{-0.3cm}
    \label{fig:bkgd}
\end{figure}

\vspace{2mm}
\noindent\textbf{Generative language model inference. } Transformers have revolutionized language models, enabling systems like BERT~\cite{devlin2018bert} to predict missing words within sentences, thus enhancing our understanding of language contexts. Yet, with the rise of applications such as ChatGPT, generative large language models have taken center stage. These models diverge from mere language understanding by engaging in autoregressive token generation -- taking a user prompt and iteratively predicting subsequent tokens until an end-of-sequence (EOS) token emerges or a predefined limit is reached, as shown in Fig.~\ref{fig:bkgd}. A key component supporting this process is the KV cache, which stores key and value vectors from previously processed tokens. This mechanism allows for subsequent tokens to be processed with attention scores computed against all prior KV vectors without the need for recomputation, enabling the LLM to efficiently generate significantly more tokens than the input prompt. As a result, the computational and carbon footprint during the inference phase is primarily driven by token generation, rather than the initial pre-filing phase to populate the input prompt's KV vectors~\cite{kwon2023efficient}. Note that in the context of this work, all LLMs we refer to are generative models. For a deeper dive into the intricacies of LLM inference, readers are encouraged to consult previous works~\cite{yu2022orca,gim2023prompt}.

\label{sec:bkgd}

\subsection{Discoveries and Opportunities}
\label{sec:motiv}

In this section, we introduce a unique discovery for generative language model inference that distinguishes this paper from all previous works and discuss how \sol{} can exploit it to save inference carbon footprint.

\begin{figure}[t]
    \centering
    \includegraphics[scale=0.47]{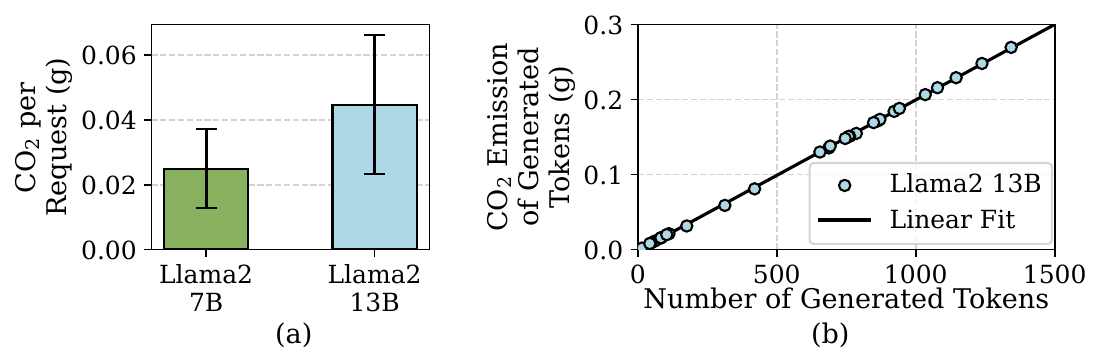}%\newline
    \vspace{1mm}
    \hrule
     \vspace{1mm}
    \caption{Two factors that impact a request's carbon footprint during LLM inference: (a) the number of model parameters and (b) the number of generated tokens. }
    % \vspace{-0.3cm}
    \label{fig:motiv1}
\end{figure}

\vspace{1.5mm}
\noindent\textbf{Takeaway 1. The carbon footprint of LLM inference depends on not only the model size but also the number of tokens generated, presenting a new opportunity to reduce carbon without reducing the model size.}

% In Fig.~\ref{fig:motiv1} (a), we show that using LLM models of different numbers of parameters (Llama2 with 7 billion and 13 billion parameters) varies the inference carbon footprint. This is the pivotal point that previous works such as INFaaS~\cite{romero2021infaas}, Clover~\cite{li2023clover}, and ALERT~\cite{wan2020alert} have exploited to explore the model parameter configurations to optimize cost, carbon, and energy. However, in the generative AI space, we find another dominating factor that did not exist in traditional ML inference systems -- the number of tokens it generates for a prompt. 

In Fig.~\ref{fig:motiv1} (a), we demonstrate how the carbon footprint of LLM inference changes with model size, showcasing examples with the Llama2 model at 7 billion and 13 billion parameters. Previous studies, such as those by INFaaS~\cite{romero2021infaas}, Clover~\cite{li2023clover}, and ALERT~\cite{wan2020alert}, have delved into optimizing model parameters to reduce costs, carbon, and energy consumption. Yet, our research uncovers a previously unexplored factor in generative AI that significantly influences carbon emissions: the number of tokens generated in response to a prompt. 

In Fig.~\ref{fig:motiv1} (b), we execute a series of input prompts on the Llama2 13B model and observe that there is a strong linear correlation between the total carbon emission and the volume of tokens generated from request. Despite initial computations to pre-fill the KV cache with key and value vectors from the input prompt, we show that \textit{the overall carbon emission of a request is largely dictated by the quantity of generated tokens}. This revelation opens up a novel pathway for optimizing the carbon efficiency of generative language model inference. Rather than downsizing the model and potentially compromising its contextual understanding capabilities, maintaining the model's size while focusing on generating fewer, more concise tokens can be the key step toward more sustainable GenAI.

\begin{figure}[t]
    \centering
    \includegraphics[scale=0.45]{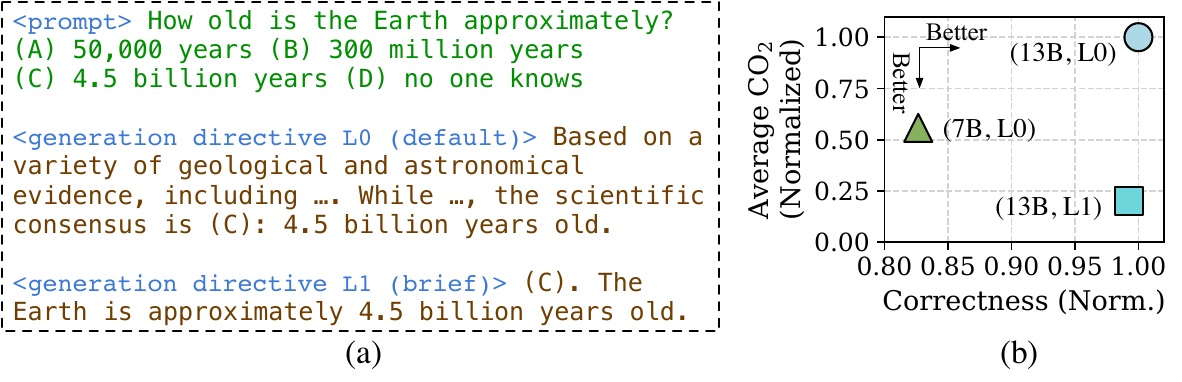}%\newline
    \vspace{1mm}
    \hrule
     \vspace{1mm}
    \caption{(a) Using generation directives can control the number of generated tokens while providing accurate responses. (b) Hosting larger models (e.g., Llama2 13B) with generation directives is better than hosting smaller models (e.g., Llama2 7B) in terms of both carbon emission and correctness.}
    % \vspace{-0.3cm}
    \label{fig:motiv2}
\end{figure}

\vspace{1.5mm}
\noindent\textbf{Takeaway 2. Incorporating generation directives into prompts can significantly reduce the carbon footprint by enabling concise yet accurate responses.} 

To control the length of token generation by a LLM, we introduce a novel concept termed "generation directive," defined as follows:
\vspace{1mm}

\begin{definition}
A \textbf{generation directive} is an instruction associated with a prompt input that dictates the manner in which a generative language model produces tokens for the prompt. Each \textbf{generation directive level} specifies a pre-defined text sequence that acts as this guiding instruction.
\end{definition} \vspace{1mm}

Similar to a compiler directive, which orchestrates the program compilation process, a generation directive strategically influences the LLM's token generation for an input prompt (details in Sec.~\ref{sec:implement}). In Fig.~\ref{fig:motiv2} (a), we show a prompt from the popular MMLU task~\cite{hendrycks2020measuring}. Without using specific directives (denoted as level L0), the Llama2 13B model defaults to generating an extensive number of tokens to elucidate the selection. However, such detailed background information may not always align with user preferences. Implementing a generation directive at level L1, designed to prompt the LLM toward brief responses, ensures both brevity and correctness. This application demonstrates a significant reduction in carbon emissions from generated tokens, as evidenced previously in Fig.~\ref{fig:motiv1} (b). Since the generation-directive-induced tokens are stored in the KV cache (Sec.~\ref{sec:bkgd}) to incur minimal additional emissions, a generation directive would significantly enhance the carbon efficiency of generative language model inference.

Fig.~\ref{fig:motiv2} (b) demonstrates that employing generation directives with a larger model (13B, L1) significantly outperforms smaller models (7B, L0) in both carbon efficiency and the accuracy of generated content. This is attributed to the larger model's superior contextual understanding, which, when combined with concise generation directives, retains its comprehensive knowledge base without unnecessary verbosity. This approach not only reduces the carbon footprint but also ensures high-quality outputs, highlighting the strategic advantage of optimizing response generation over simply reducing model size.

\begin{figure}[t]
    \centering
    \includegraphics[scale=0.54]{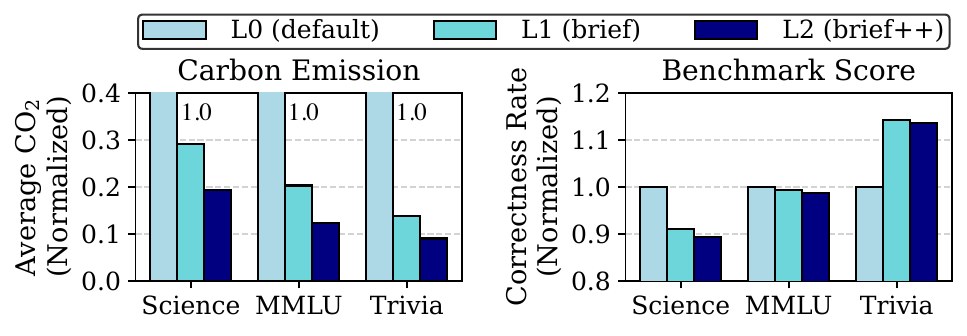}%\newline
    \vspace{1mm}
    \hrule
     \vspace{1mm}
    \caption{Applying generation directives across different applications reveals variability in sensitivity to these directives, impacting both carbon emissions and the accuracy of the generated content.}
    % \vspace{-0.3cm}
    \label{fig:motiv3}
\end{figure}

\begin{figure*}[ht]
    \centering
    \includegraphics[scale=0.45]{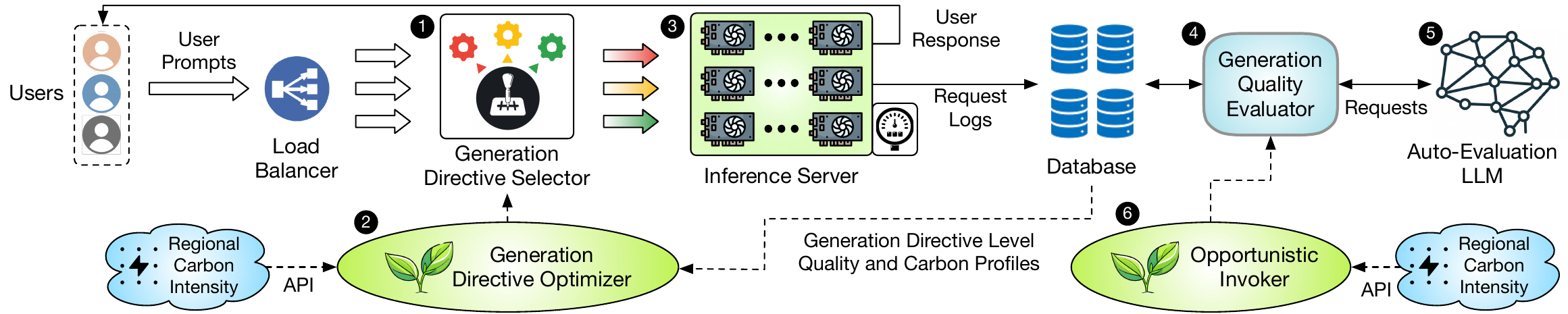}  
    \vspace{.1cm}
    \hrule  
    \vspace{1mm}
    \caption{System Design Overview of \sol{}.}
    \vspace{-4mm}
    \label{fig:desi0_overview}
\end{figure*}

\vspace{1.5mm}
\noindent\textbf{Takeaway 3: The impact of employing generation directives on carbon emissions and accuracy differs across user prompts, presenting a systemic challenge in optimally utilizing these directives, particularly in the context of fluctuating carbon intensity.} 

We have shown the effectiveness of generation directives on the MMLU task but in reality, the user can submit all kinds of prompts. In Fig.~\ref{fig:motiv3}, we show the impacts of different generation directives (L0, L1, L2) on different tasks including science knowledge~\cite{lu2022learn} and trivia knowledge~\cite{joshi2017triviaqa}. We can observe that both the amount of carbon emission and the generation's correctness rate vary with the task. The findings indicate that while directives promoting concise responses may decrease accuracy in complex, multi-step reasoning tasks, they could enhance it in tasks where answers are directly inferable from the prompt or learned context. Therefore, a system configuring the generation directives must have a generation quality evaluator (Sec.~\ref{sec:design_overview}) to provide feedback when the user prompts' sensitivity to directive levels varies over time. 

In addition, all the CO$_2$ emissions in this section are shown with a constant carbon intensity of 100 gCO$_2$/kWh and power usage effectiveness (PUE) of 1.2, while in the real world, the carbon intensity changes all the time from the varying energy source mixture~\cite{acun2023carbon,li2023toward}. Responding to these challenges, we design \sol{}, a generative language model inference framework that takes advantage of generation directives to dynamically optimize the carbon footprint while guaranteeing high-quality generations.

\section{Design}
\label{sec:design}

\subsection{System Overview}
\label{sec:design_overview}

\sol{} is designed as the first carbon-aware generative language model inference framework, utilizing token generation directives to minimize carbon footprint while ensuring high-quality content generation. Fig.~\ref{fig:desi0_overview} shows a brief design overview of \sol{}. Once the users prompts are assigned to an inference server by the load balancer, the prompts need to be tokenized into numerical representations. In this phase, a generation directive selector \blackcircled{1} assigns a directive level to each prompt, integrating it into the tokenized input. The policy to assign the directive levels is determined by the \sol{}'s token generation directive optimizer \blackcircled{2} (Sec.~\ref{sec:desi_formulation}), which is based on the current electricity grid's carbon intensity and the token generation quality and carbon feedback. 

To retrieve the local carbon intensity, we can access third-party API endpoints such as Electricity Maps~\cite{electricitymap}. To enable inference carbon feedback, we can monitor the datacenter PUE and device energy with tools such as \texttt{nvidia-smi} to record the GPU power and processing time of requests and save the logs to the database. However, obtaining the token generation quality feedback is a different process from the above metrics. 
After autoregressive inference concludes on the inference server \blackcircled{3}, the generated tokens are detokenized and sent back to the user clients, while simultaneously, the request and node monitoring logs are archived in the database. A generation quality evaluator \blackcircled{4} then extracts a sample of prompts from the database, generates responses for each at all generation directive levels, and identifies the directive level that yields the best response for each request. 

Determining the optimal level for quality generation presents a challenge due to the subjective nature of preference and the absence of a definitive best response for user prompts, making manual evaluation by humans impractical. Following a methodology from recent research~\cite{dubois2024alpacafarm}, an LLM-based automatic evaluator, rather than human evaluators, is employed to assess generation feedback, aligning with common academic and industry practices~\cite{liu-etal-2023-g,bai2024benchmarking,mistral_prompting}. This evaluator consults an auto-evaluation LLM \blackcircled{5} to gauge its preference on the responses, logging these preferences back into the database. The whole process happens offline, and since the evaluation process also incurs carbon emission, \sol{}'s opportunistic evaluation invoker \blackcircled{6} (Sec.~\ref{sec:desi_evaluator}) ensures the evaluations are carried out only as necessary and during periods of low carbon intensity.

Next, we explore the foundational mechanisms of \sol{}, focusing on its strategic use of generation directives via the token generation directive optimizer to both reduce the carbon footprint and ensure the generation of high-quality content.

\subsection{Generation Directive Optimizer}
\label{sec:desi_formulation}
% selecting the optimization variable
% - Selecting the lvl for each prompt is inefficient
% - Selecting the global lvl selection probability is better

Section~\ref{sec:motiv} illustrates that while employing generation directives to reduce token output in the autoregressive process is beneficial for lowering carbon emissions, it poses a risk to content quality. Two key external factors further complicate this balance: the regional carbon intensity powering the datacenter, which directly affects the efficacy of carbon savings, and the nature of user prompts, which influences the impact of generation directives on both emissions and content quality. To address these challenges, \sol{}'s optimizer is designed to dynamically adjust to fluctuations in carbon intensity and the variability of user prompt tasks. In scenarios of low carbon intensity, \sol{} prioritizes directives that enhance content quality, leveraging the reduced carbon cost of generating new tokens. Conversely, under high carbon intensity, it opts for directives that may slightly compromise quality but significantly reduce emissions. This strategic approach underpins the mathematical formulation of the \sol{} optimizer, ensuring that it aligns with the dual objectives of environmental sustainability and content quality. \vspace{2mm}

\noindent\textbf{Optimization variable. } The core challenge lies in selecting the optimal generation directive for each user request to minimize carbon emissions while ensuring satisfactory generation quality. However, optimizing directive levels on a per-prompt basis introduces several practical complications: (i) Dimensionality challenge: optimizing for a per-prompt basis brings up the dimensionality challenge as the number of dimensions equals the number of requests at each optimization step.
(ii) Computational overhead: the optimization is in the critical path before the autoregressive inference starts and could introduce significant overhead as tens to hundreds of new requests can arrive every second. Solving a high-dimension optimization problem would require significant compute cycles that delay the time to first token (TTFT).
(iii) Predictability issues: anticipating the specific impact of each directive level on carbon emissions and content quality for individual prompts is challenging. While general trends can be inferred from historical data, the unique nature of each prompt means accurate predictions are only feasible post-inference.

Considering the outlined design challenges, \sol{} adopts a system-level optimization strategy for generation directive levels, rather than an impractical per-prompt optimization. It achieves this by determining the probability of selecting each directive level for any user prompt received. Let's denote $n$ as the total number of available generation directive levels. The optimization variable, represented as $\mathbf{x} = [x_0, x_1, \ldots, x_{n-1}]^T$, defines $x_i \in [0,1]$ as the probability of applying the $i$-th directive level to any prompt, with $x_0$ representing the baseline directive L0 (indicating no directive). To ensure every prompt receives a directive level, the condition $\sum_{i=0}^{n-1} x_i = 1$ must be satisfied. This system-wide probabilistic approach to directive selection, while not optimizing for individual prompts, is shown to achieve carbon savings close to those of an impractical per-prompt Oracle optimizer, as detailed in Sec.~\ref{sec:eval}. \vspace{2mm}

\noindent\textbf{Objective function. } The primary goal of \sol{} is to minimize the carbon footprint associated with each inference request. The objective function, $f(\mathbf{x})$, encapsulates the expected carbon footprint of an inference request, detailed in Eq.\ref{eq:bkgd} (Sec.~\ref{sec:bkgd}). It incorporates: (i) the current regional carbon intensity ($k_0$ in $gCO_2/kWh$), obtained via API; (ii) the prorated per-second embodied carbon of the inference hardware through its device lifetime ($k_1$ in $gCO_2/s$); and (iii) the profiles of energy consumption ($\mathbf{e}$) and processing time ($\mathbf{p}$) for requests employing various generation directive levels. The vectors $\mathbf{e} = [e_0, e_1, \ldots, e_{n-1}]^T$ and $\mathbf{p} = [p_0, p_1, \ldots, p_{n-1}]^T$ represent the average energy (in kWh) and processing time (in seconds), respectively, for recent requests at each directive level, retrievable from the database. Following Eq.~\ref{eq:bkgd}, the formula for calculating the expected carbon emissions of an inference request is thus given by:
\begin{align}
\label{eq:desi_obj}
f(\mathbf{x}) = k_0 \cdot \mathbf{e}^T \mathbf{x} + k_1 \cdot \mathbf{p}^T \mathbf{x},
\end{align}
where $\mathbf{x}$ denotes the probabilities of selecting each directive level across all user prompts. \vspace{2mm}

\noindent\textbf{Generation quality constraints. } The last piece of information the optimizer needs is quality feedback. The generation quality evaluator reports the auto-evaluation LLM's preference on which directive level is the best for all sampled requests. Let $\mathbf{q} = [q_0, q_1, \ldots, q_{n-1}]^T$ where $q_i\in [0,1]$ denote the preference rate of each directive level reported by the evaluator. For example, if $\mathbf{q} = [0.5, 0.3, 0.2]^T$, it means 50\% of the time, the auto-evaluator prefers the response generated using directive L0, 30\% of the time by L1 and 20\% of the time by L2. We can denote the expected generation quality as $\mathbf{q}^T\mathbf{x}$. During the optimization, we need to make sure the preference rate does not deviate beyond a threshold of $\xi \in [0,1]$ away from the $q_0$ generation baseline using directive $L0$. In addition, \sol{} designs the actual quality deviation from $q_0$ to vary based on the current carbon intensity -- when the carbon intensity is low, the constraint should be more strictly enforced (deviation closer to 0) since renewable energy is abundant in the grid to support high-quality generation, and vice versa, during high carbon intensity periods, the deviation should be closer to $\xi$. This can be formulated as an inequality constraint:
\begin{align}
\label{eq:desi_quality}    
\mathbf{q}^T\mathbf{x} \geq (1 - \frac{k_0-k_0^\text{min}}{k_0^\text{max}-k_0^\text{min}} \cdot \xi) \cdot q_0
\end{align}
where $k_0^\text{min}$ and $k_0^{\text{max}}$ are the known historical minimum and maximum carbon intensities, respectively. The parameter $\xi$, adjustable according to system requirements, facilitates a balance between carbon footprint and content quality. For \sol{}'s evaluation (detailed in Sec.~\ref{sec:eval}), we set $\xi$ to 0.1. This setting dictates that no matter how high the carbon intensity is, the system must select directive levels that ensure the auto-evaluation LLM's preference for generated content remains at least 90\% as favorable as it would be using the baseline directive L0. \vspace{2mm}

\noindent\textbf{Problem formulation. } We can construct the overall optimization problem using the objective function in Eq.~\ref{eq:desi_obj} and the constraint in Eq.~\ref{eq:desi_quality} along with other inherent constraints of $\mathbf{x}$. For simplicity, we replace the right-hand side of Eq.~\ref{eq:desi_quality} with scalar $q_{lb}$ to represent the quality lower bound. We have
\begin{align}
    \label{eq:desi_opt0}
    &\min_{\mathbf{x}\in\mathbb{R}^n}
    \begin{aligned}[t]
      & f(\mathbf{x})
    \end{aligned} \\    
    & \label{eq:desi_opt1}\text{s.t. } \;\mathbf{q}^T\mathbf{x} \geq q_{lb}, \\
    & \label{eq:desi_opt2} \;\;\;\;\;\;\;\forall i,\; 0 \leq x_i \leq 1, \\
    & \label{eq:desi_opt3} \;\;\;\;\;\; \sum_{i=0}^{n-1} x_i = 1
\end{align}
where the inequality constraint Eq.~\ref{eq:desi_opt2} indicates that the probability of each level is within the range of 0 to 1, and the equality constraint Eq.~\ref{eq:desi_opt3} indicates that all probabilities sum to 1.
We can observe that the objective function Eq.~\ref{eq:desi_obj} is linear because both $\mathbf{e}^T$ and $\mathbf{p}^T$ are constants to the optimization variable $\bm{x}$. In addition, all the constraints in Eq.~\ref{eq:desi_opt1},~\ref{eq:desi_opt2} and~\ref{eq:desi_opt3} are all linear to $\mathbf{x}$. Therefore, we have mapped the optimal generation directive level configuration problem to a linear programming problem and we can use the HiGHS dual simplex solver~\cite{huangfu2018parallelizing} to find the optimal solution for $\mathbf{x}$.

\subsection{Opportunistic Offline Quality Assessment}
\label{sec:desi_evaluator}

In Eq.~\ref{eq:desi_opt1}, \sol{} relies on the $\mathbf{q}^T$ vector to impose the quality constraint. As a carbon-friendly generative language model inference framework, \sol{} not only cares about the carbon footprint of the inference server but also the quality evaluation process, especially when the auto-evaluation LLM can have $> 10\times$ number of parameters than the inference model (e.g., the GPT-4 model with a Mixture-of-Experts architecture is estimated to have 220B parameters per expert~\cite{gpt4wiki} comparing against the Llama2 13B model). Note that the quality evaluation is not in the critical path of online inference serving because it is not latency-critical and thus can be done offline in a different server as an application decoupled from inference. 

Consequently, \sol{} adopts an opportunistic approach to performing generation quality evaluations, triggering them based on specific carbon intensity thresholds of the evaluation server. This strategy is informed by the premise that (i) access to the auto-evaluation LLM might be facilitated exclusively via third-party APIs, such as OpenAI's API, and (ii) the volume of evaluation requests remains constant across cycles, as detailed in Sec.~\ref{sec:implement}. This method ensures that \sol{}'s carbon footprint overhead from the quality evaluation is minimized.

When deciding on whether to evaluate at the current time $t$, it's critical to weigh the carbon intensity of the LLM at the current moment, denoted as $k_2^{(t)}$, against the time elapsed since the last evaluation at $t_0$. Direct and frequent evaluations can lead to unnecessary carbon emissions without significant benefit, whereas delayed evaluations can undermine the optimizer's reliability, as the $\mathbf{q}^T$ vector becomes outdated (Sec.~\ref{sec:desi_formulation}). To mitigate these issues, we first enforce a grace period to ensure the evaluation does not occur too frequently, then introduce an urgency multiplier to the carbon intensity to capture the increasing need for re-evaluation as time progresses. The urgency-adjusted carbon intensity $k_2'^{(t)}$ can be expressed as
\begin{align}
\label{eq:desi_evaluator}    
k_2'^{(t)} = e^{-\beta(t-t_0)}\cdot k_2^{(t)}
\end{align}

\begin{figure}[t]
    \centering
    \includegraphics[scale=0.54]{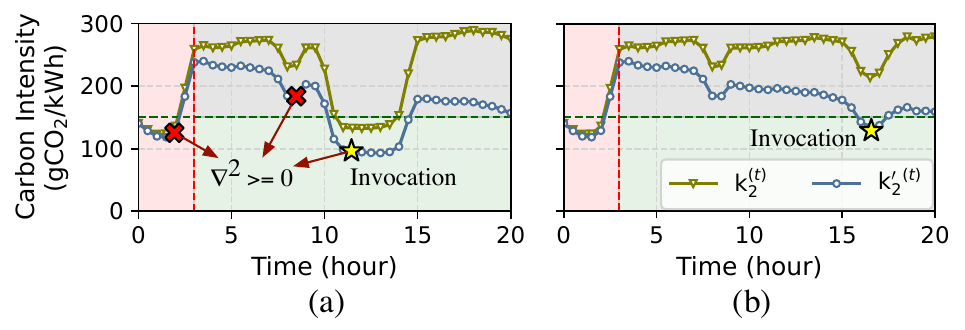}%\newline
    \vspace{1mm}
    \hrule
     \vspace{1mm}
    \caption{Process in selecting the generation quality evaluation opportunity (marked as a golden star). The grace period is depicted by a red area, while the carbon intensity threshold is indicated by a green horizontal line. (a) For an offline evaluation to be deemed appropriate, the urgency-adjusted carbon intensity, $k_2'^{(t)}$, must fall within the green zone. Instances marked by two red crosses, despite showing a positive second-order derivative, do not qualify for evaluation due to their positioning outside the eligible range. (b) Even if carbon intensity stays high all the time, the increasing evaluation urgency ensures that offline evaluation always occurs.}
    % \vspace{-0.3cm}
    \label{fig:desi_evaluator}
\end{figure}

The urgency parameter, $\beta$, determines the rate at which the evaluation interval incurs penalties over time, ensuring that the value of immediate evaluation -- offering a timely update to the $\mathbf{q}^T$ vector in Sec.\ref{sec:desi_formulation} -- is weighed against waiting for potentially lower future carbon intensities. Setting $\beta$ to 0.028, for instance, halves the urgency-adjusted carbon intensity, $k_2'^{(t)}$, relative to the actual carbon intensity, $k_2^{(t)}$, after a 24-hour lapse without evaluation. An offline evaluation is initiated under three conditions: (i) $t_s$ represents a local minimum for $k_2'^{(t)}$, indicating a positive second-order derivative at that point; (ii) a grace period has elapsed since the last evaluation; (iii) the urgency-adjusted carbon intensity at $t_s$, $k_2'^{(t_s)}$, falls below a predefined threshold, such as 50\% of the historical maximum carbon intensity. This evaluative mechanism, illustrated in Fig.~\ref{fig:desi_evaluator}, highlights moments of evaluation marked by stars in two different cases, underlining the proactive approach of \sol{} in scheduling evaluations with consideration for both carbon intensity and the need for timely quality feedback. 
% Note that in the figure, at each timestamp, all the future carbon intensities are unknown.

\subsection{Miscellaneous Design Considerations}
\noindent\textbf{Role of auto-evaluation LLM. } The auto-evaluation LLM, boasting orders of magnitude more parameters than the inference model, might seem like an ideal choice for processing user prompts. However, utilizing a giant model like GPT4, with its estimated 1.76 trillion parameters~\cite{gpt4wiki}, entails considerable development, training, and deployment resources, making it impractical for most organizations due to high costs and environmental impact. Also, directly serving millions of user prompts on such a model incurs significantly more carbon emissions than a model with billions of parameters. Therefore, for most cases, it is better to fine-tune an open-sourced model like Llama2 to tailor to the user targets and use third-party LLMs like GPT4 for occasional quality feedback. 

There may be instances where the auto-evaluator's preferences diverge from an individual user's expectations, as users might have varying inclinations toward the conciseness or detail of responses. In such cases, the inference service could proactively notify users when responses are condensed due to elevated carbon intensity levels, subsequently inquiring about their preference for more detailed answers. Should a user client express a preference for depth, \sol{} can then specifically mark this preference by applying the baseline directive, L0, to all their future prompts, ensuring tailored responses that align more closely with their expectations.

\vspace{2mm}
\noindent\textbf{Number of evaluation samples. } According to the sample size theory in~\cite{charan2013calculate}, at each quality evaluation, with a confidence level of 95\%, if we sample 500 user prompts, the maximum margin of error is only 4.4\%. Therefore, we use a fixed-sized 500 request samples to provide generation quality feedback. 
This fixed sample size of 500 is chosen for generating quality feedback within \sol{}, considering its minimal impact relative to the total volume of prompts processed during the evaluation period. Consequently, the carbon emissions associated with these evaluations are deemed negligible and are not factored into the carbon footprint reduction strategy detailed in Sec.~\ref{sec:desi_formulation}.

\subsection{Implementation}
\label{sec:implement}

\begin{figure}[t]
    \centering
    \includegraphics[scale=0.42]{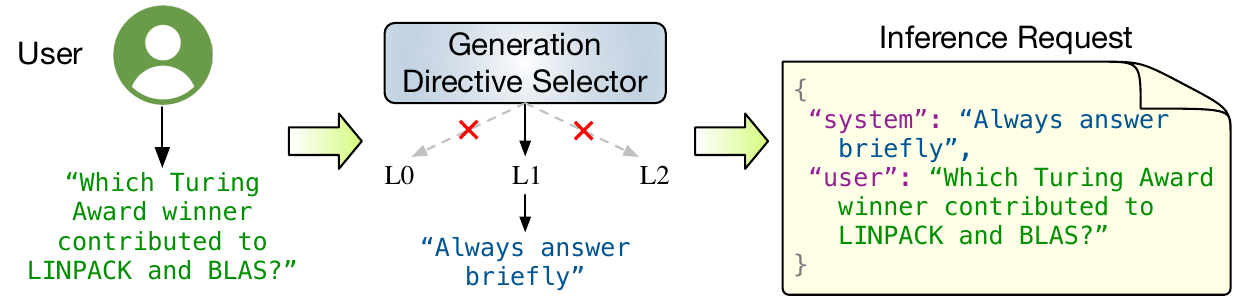}%\newline
    \vspace{1mm}
    \hrule
    \vspace{1mm}
    \caption{\sol{} implements generation directive level assignment as LLM system prompts.}
    % \vspace{-0.3cm}
    \label{fig:design2}
\end{figure}

\noindent\textbf{Applying generation directive levels. } The inference service provider specifies the number of directive levels and the actual directive to apply for each level. \sol{} implements the generation directives as the system prompt alongside the user prompt, as the system prompt is widely accepted as a prompting format compatible with leading AI platforms like OpenAI ChatML~\cite{chatml}, Llama~\cite{llama}, Anthropic Claude~\cite{anthropic}, MistralAI~\cite{mistralai}, etc. 
Figure~\ref{fig:design2} illustrates \sol{}'s method of incorporating a specific directive, such as the text from level L1, directly into the inference request as a system prompt. When a system prompt already exists within a user prompt, \sol{} precedes it with the selected generation directive, ensuring seamless integration.
\vspace{2mm}

% Fig.~\ref{fig:design2} describes the process of \sol{} applying a specific directive on a user prompt example, as the pre-defined directive text from L1 is included as a system prompt in the inference request. If the user prompt always has a system prompt, \sol{} will instead prepend the generation directive to the existing system prompt. 

\noindent\textbf{Inference server and monitoring. } \sol{} seamlessly integrates with existing inference server setups by processing system prompts together with user prompts, avoiding the need for infrastructure alterations. Mirroring industry-standard LLM inference practices, the server incorporates vLLM~\cite{kwon2023efficient} for its high-throughput and efficient KV cache management and utilizes FlashAttention~\cite{dao2023flashattention} to streamline self-attention computations at the CUDA kernel level. To accurately log execution metrics as outlined in Eq.~\ref{eq:desi_obj}, the CarbonTracker~\cite{anthony2020carbontracker} package has been adapted to monitor each inference processing node, facilitating the calculation of $\mathbf{e}^T$ and $\mathbf{p}^T$ vectors essential for optimizing \sol{}'s operation.
\vspace{2mm}

\begin{figure}[t]
    \centering
    \includegraphics[scale=0.355]{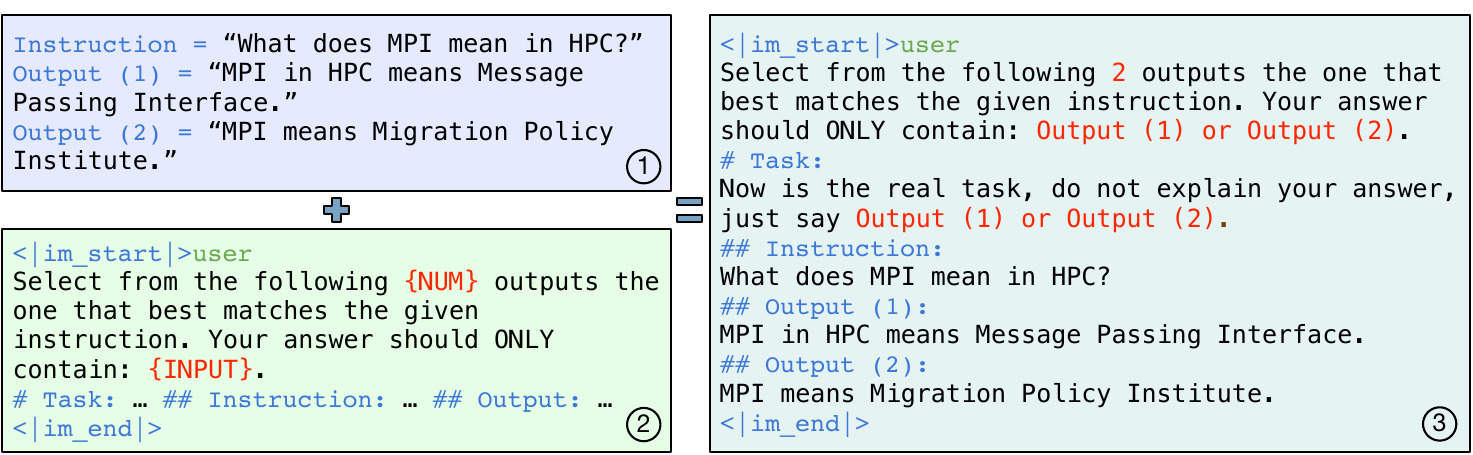}%\newline
    \vspace{1mm}
    \hrule
    \vspace{1mm}
    \caption{A simplified example of \sol{}'s quality evaluation query. Box 1 represents the instructions and outputs generated using different directives, box 2 represents the quality evaluator template, and box 3 represents the query in ChatML~\cite{chatml} format to the auto-evaluation LLM.}
    % \vspace{-0.3cm}
    \label{fig:design3}
\end{figure}

\noindent\textbf{Automatic quality evaluation. } We extend the AplacaEval~\cite{alpaca_eval} project to build \sol{}'s quality evaluator. Specifically, we generalized the auto-annotator to be able to query the auto-evaluation LLM to select the best one from an arbitrary number of generations, each corresponding to a specific generation directive level. We also implemented shuffling of the directive-guided generations to remove position bias in the query. The evaluator is meticulously implemented to prompt the auto-evaluation LLM to generate minimal tokens -- just enough to identify the preferred output prior to the EOS token. This design is both carbon-efficient and cost-effective as commercial LLMs charge based on the number of tokens generated. A simplified example is shown in Fig.~\ref{fig:design3} where when we send the query to auto-evaluation LLM, it will generate ``Output (1)'' as the preferred output. We have manually examined the preference of several auto-evaluation LLMs (GPT-4, GPT-4 Turbo, GPT-3.5 Turbo) and confirm that the evaluator accurately identifies the correct response in over 97\% of cases.

% both the generation directive and evaluator template cannot be too long, otherwise introduces too many more tokens

% both evaluation opportunity and prompt design are low carbon

% explain how to add system prompt, HTTP json

\section{Methodology}
\label{sec:method}

\noindent\textbf{Experiment setup. } 
% LLAMA2, GPU node, directive levels
The experiments are carried out on a testbed comprising two nodes, each equipped with two NVIDIA A100 40GB Tensor Core GPUs and two AMD EPYC 7542 CPUs. The Llama2 13B model, a prominent large language model released by Meta~\cite{touvron2023llama}, is utilized to establish the inference server, with each GPU hosting a model instance within its 40GB HBM memory. To assess \sol{}'s efficiency, three levels of generation directives are implemented: L0 as the default baseline with no directives, L1 for ``brief" generation, and L2 for ``very brief" generation. GPT-4, accessed via the OpenAI API, serves as the auto-evaluation LLM for offline quality assessments.

\begin{table}
\centering
\caption{Language modeling tasks to evaluate \sol{}.}
% \vspace{-4mm}
\scalebox{0.9}{
\begin{tabular}{ccc}%P{6.7cm}P{1.7cm}P{1.7cm}P{1.7cm}} 
%  \hline
%  \multicolumn{1}{|c|}{\textbf{Category}} & \multicolumn{1}{|c|}{\textbf{Instance types}} & \multicolumn{1}{|c|}{\textbf{Size}} \\ [0.5ex] 
\toprule
\textbf{Dataset} & \textbf{Description} & \textbf{Task} \\ 
\midrule
\midrule
Alpaca~\cite{taori2023alpaca} & \makecell[c]{Instructions generated by\\OpenAI's \texttt{text-davinci-003}} & Instruction tuning \\
\midrule
GSM8K~\cite{cobbe2021training} & \makecell[c]{Grade school\\math problems} & \makecell[c]{Arithmetic and\\multi-step reasoning} \\
\midrule
MMLU~\cite{hendrycks2020measuring} & \makecell[c]{Massive multitask\\language understanding} & \makecell[c]{Multiple-choice\\questions} \\
\midrule
\makecell[c]{Natural\\Questions~\cite{kwiatkowski2019natural}} & \makecell[c]{Real-user questions\\from Google} &  \makecell[c]{Question\\answering} \\
\midrule
ScienceQA~\cite{lu2022learn} & \makecell[c]{School science subjects\\(e.g., Biology/Physics/Chemistry)} &  \makecell[c]{Multiple-choice\\science questions} \\
\midrule
TriviaQA~\cite{joshi2017triviaqa} & \makecell[c]{Trivia questions collected\\by trivia enthusiasts} & \makecell[c]{Reading\\comprehension} \\
\bottomrule
\end{tabular}}
\vspace{-1mm}
\label{table:method1}
\end{table}

\sol{} is evaluated using six diverse language modeling datasets, detailed in Table~\ref{table:method1}. These datasets span various fields and applications, serving as critical benchmarks in performance evaluations for leading large language models, including Llama~\cite{touvron2023llama}, Claude~\cite{claude3}, Mixtral~\cite{jiang2024mixtral}, GPT~\cite{achiam2023gpt}, and Gemini~\cite{team2023gemini}. To simulate realistic user prompts for the inference server, the construction of tasks is guided by user request patterns from the Alibaba Platform for AI trace~\cite{weng2022mlaas}, ensuring the evaluation comprehensively represents the workload encountered in practical scenarios.

\begin{table}
\centering
\caption{Different geographical regions and their minimum and maximum annual carbon intensity.}
% \vspace{-4mm}
\scalebox{0.88}{
\begin{tabular}{cccc}%P{6.7cm}P{1.7cm}P{1.7cm}P{1.7cm}} 
%  \hline
%  \multicolumn{1}{|c|}{\textbf{Category}} & \multicolumn{1}{|c|}{\textbf{Instance types}} & \multicolumn{1}{|c|}{\textbf{Size}} \\ [0.5ex] 
\toprule
\textbf{Region} & \textbf{abbr.} & \textbf{Operator} & \textbf{Annual Min/Max} \\ 
\midrule
\midrule
Texas (US) & TX & \makecell[c]{Electric Reliability
\\Council of Texas (ERCOT)} & 124 / 494 (gCO$_2$/kWh) \\
\midrule
California (US) & CA & \makecell[c]{California Independent\\System Operator (CISO)} & 55 / 331 (gCO$_2$/kWh)\\
\midrule
South Australia & SA & \makecell[c]{Australian Energy\\Market Operator (AEMO)} & 10 / 526 (gCO$_2$/kWh)\\
\midrule
Netherland & NL & TenneT & 23 / 463 (gCO$_2$/kWh)\\
\midrule
Great Britain & GB & \makecell[c]{National Grid Electricity\\System Operator (ESO)} & 24 / 282 (gCO$_2$/kWh)\\

\bottomrule
\end{tabular}}
\vspace{-2mm}
\label{table:method2}
\end{table}

The evaluation of \sol{} extends across five grid operation regions in various countries, as described in Table~\ref{table:method2}. Given the variability in carbon intensity by region, this diversity enables a comprehensive assessment of \sol{}'s performance in differing environmental contexts. The study uses carbon intensity data from February, June, and October of 2023, sourced from Electricity Maps~\cite{electricitymap} at hourly intervals, to gauge \sol{}'s adaptability to fluctuating carbon intensity levels across these regions. Despite the offline evaluation LLM not being sensitive to latency and thus not requiring proximity to users -- allowing it to be located in any global data center with the lowest carbon footprint -- for a more cautious approach, we assume it resides in the same region as the inference server. \vspace{2mm}

\noindent\textbf{Competing schemes. } \sol{} is evaluated alongside five distinct strategies, detailed as follows:

\textbf{\base{}. } This is the baseline strategy that represents a vanilla LLM inference system, it does not explore the opportunity of generation directives discussed in Sec.~\ref{sec:motiv}. 

\textbf{\cooopt{}. } This represents a scheme that aggressively minimizes CO$_2$ emissions without considering the generation quality. Specifically, it will always use the generation directive level that yields the lowest carbon footprint for all prompts.

\textbf{\modelopt{}. } This scheme is an implementation of the idea to adjust the underlying model parameters to achieve optimization goals from previous works~\cite{romero2021infaas,li2023clover,wan2020alert}. Unaware of the generation directives, this scheme uses inference model variants (i.e., Llama2 7B and 13B) as optimization variables since model variants also introduce the trade-offs between carbon and generation quality. The scheme represents the optimal model variant selection for the user prompts.

\textbf{\static{}. } This is a static version of \sol{}, 
applying a single, month-long optimal generation directive configuration identified through offline analysis, without dynamic adjustments based on real-time carbon intensity and generation feedback. The best static configuration is determined by sweeping the possible static configurations. 

\textbf{\oracle{}. } This is an impractical scheme based on oracle information. It assumes the inference carbon emission on every generation directive level is known ahead of time for all user prompts, and knows the exact generation quality feedback for future prompts instead of relying on sampling. It does not suffer from any profiling overheads and sampling inaccuracies. \vspace{2mm}

\noindent\textbf{Metrics. } The evaluation of \sol{} centers on two primary metrics: the carbon footprint associated with each inference request and the quality of the content it generates. The carbon footprint metric accounts for the CO$_2$ emissions associated with each inference, averaged for comparison against the default operation represented by \base{}. The generation quality is measured from the auto-evaluation LLM's preference, normalized against \base{}'s performance. For instance, if the auto-evaluator shows a preference for \sol{}'s responses 48\% of the time versus 52\% for \base{}, \sol{}'s normalized generation preference score would be 92.3\%.

Next, we thoroughly evaluate how \sol{} can contribute to sustainable GenAI using its directive-guided LLM inference.
\section{Evaluation}
\label{sec:eval}

\subsection{Effectiveness of \sol{}}
\begin{figure}[t]
    \centering
    \includegraphics[scale=0.5]{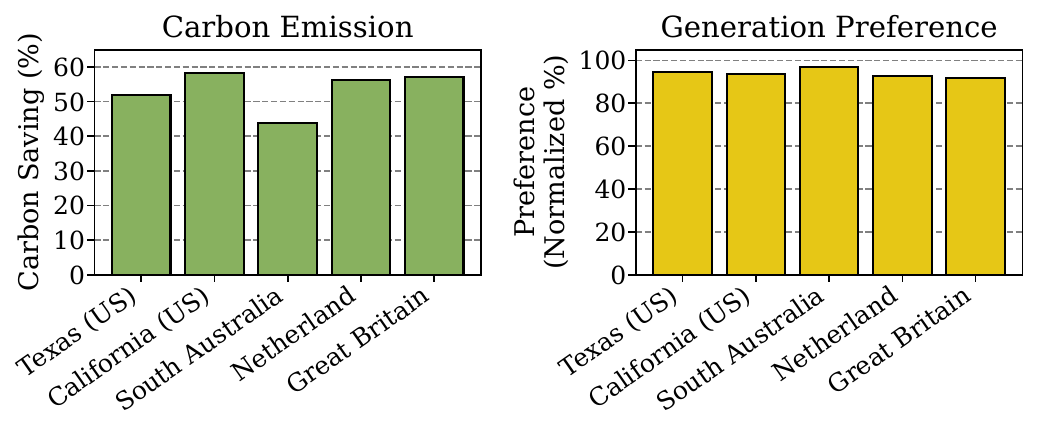}%\newline
    \vspace{1mm}
    \hrule
     \vspace{1mm}
    \caption{\sol{} archives significant carbon savings while preserving generation quality across all geographical regions.}
    % \vspace{-0.3cm}
    \label{fig:eval0}
\end{figure}

\textit{\sol{} consistently achieves substantial carbon savings while adhering to generation quality standards in diverse geographical regions.} According to Fig.~\ref{fig:eval0}, \sol{}'s application of optimized generation directives can reduce carbon emissions by up to 60\%. With a preference deviation ($\xi=0.1$) set from the baseline in Eq.~\ref{eq:desi_quality}, generation preferences across all regions remain above the 90\% mark, notably reaching over 95\% in South Australia alongside a carbon saving exceeding 40\%. From an inference service provider perspective, according to a recent survey~\cite{de2023growing}, deploying OpenAI's ChatGPT service necessitates around 29K NVIDIA A100 GPUs, equating to an energy consumption of 564 MWh daily. In the Azure West US region of California~\cite{azure}, this translates to monthly CO$_2$ emissions of 3,266 tonnes. Adopting \sol{} could result in a monthly carbon reduction of 1,903 tonnes—equivalent to offsetting the carbon footprint of flying 6,358 passengers from New York to London~\cite{flight}.

% \textit{\sol{} consistently delivers significant carbon savings while meeting generation quality constraints across all geographical regions.} From Fig.~\ref{fig:eval0}, we observe that using \sol{}-optimized generation directives can save the inference carbon emission by up to 60\%. We have set $\xi=0.1$ as the preference deviation from baseline generation in Eq.~\ref{eq:desi_quality}, and as we observe from the generation preference in Fig.~\ref{fig:eval0}, all regions have their normalized preference above 90\% while the South Australia region has above 95\% normalized preference with more than 40\% carbon emission saved. To give a more concrete view, according to a recent survey~\cite{de2023growing}, OpenAI required about 29K NVIDIA A100 GPUs to deploy the ChatGPT service, implying an energy demand of 564 MWh per day. Assuming it is deployed in the Azure West US region in California~\cite{azure}, the inference server would emit 3,266 tonnes of CO$_2$ per month. If we can deploy \sol{} to reduce carbon emissions, this translates to 1,903 tons of CO$_2$ saved per month, equivalent to the carbon emission in transporting 6,358 passengers from New York to London by flight~\cite{flight}.
%193 annual average in CA
%299.30	per passenger

\begin{figure}[t]
    \centering
    \includegraphics[scale=0.45]{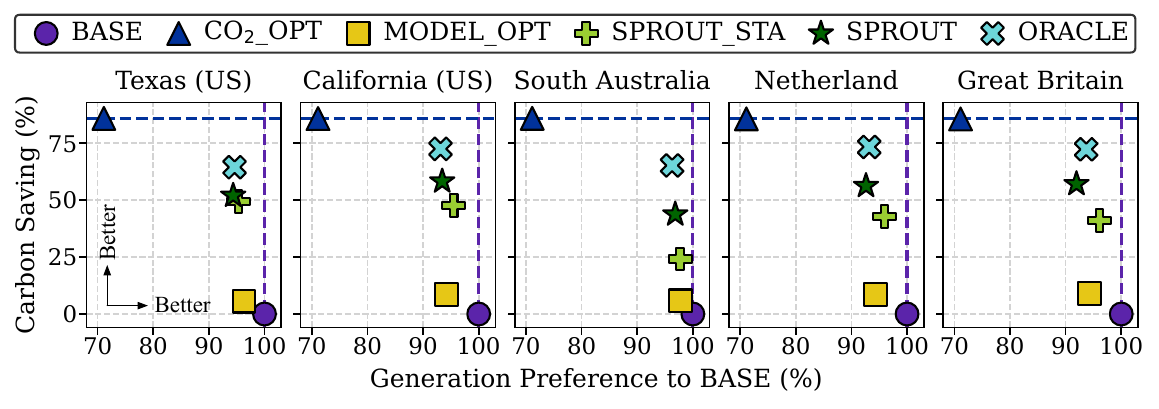}%\newline
    \vspace{1mm}
    \hrule
     \vspace{1mm}
    \caption{\sol{} excels when competing against competitive strategies and is closest to \oracle}
    % \vspace{-0.3cm}
    \label{fig:eval1}
\end{figure}

\textit{\sol{} surpasses competing methods, closely aligning with the \oracle{} standard.} Fig.~\ref{fig:eval1} illustrates \sol{}'s performance against competing strategies outlined in Sec.~\ref{sec:method}, showcasing its proximity to the ideal \oracle{} in both carbon savings and normalized generation preference across all regions. Here, vertical lines denote the upper bound of generation preference in our evaluation, while horizontal lines indicate the upper bound of carbon savings. Unlike \cooopt{}, which prioritizes carbon reduction at the expense of generation quality, \sol{} maintains a balance closer to \base{}. While \modelopt{}, \static{}, and \sol{} exhibit similar preferences, \modelopt{} falls short in carbon savings, highlighting the limitations of optimizing solely based on inference model variants~\cite{romero2021infaas,li2023clover,wan2020alert}. In contrast to its static version \static{}, \sol{} demonstrates that its dynamic approach to generation directives yields results nearer to the \oracle{} benchmark, underscoring the effectiveness of adaptive configurations.

\subsection{Mechanisms behind \sol{}'s Effectiveness}

\begin{figure}[t]
    \centering
    \includegraphics[scale=0.49]{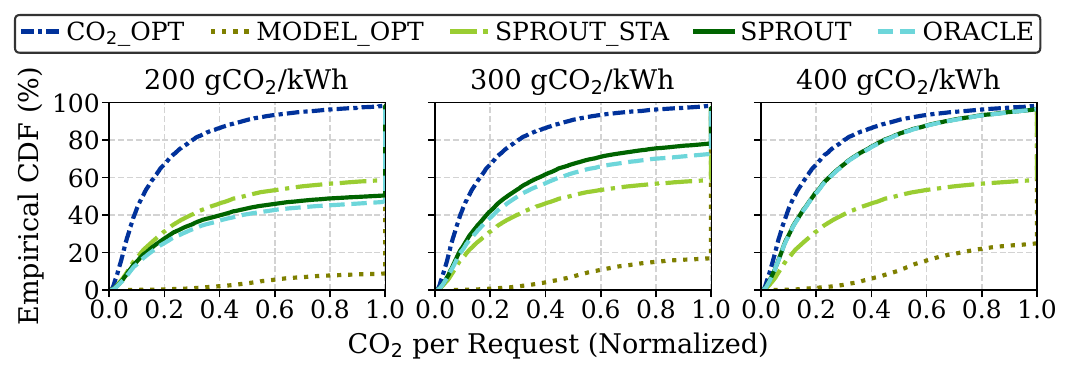}%\newline
    \vspace{1mm}
    \hrule
     \vspace{1mm}
    \caption{Cumulative distribution function of per-request carbon emission normalized to \base{} when the environmental carbon intensity varies.}
    % \vspace{-0.3cm}
    \label{fig:eval2}
\end{figure}

Next, the inference carbon footprint is analyzed from the perspective of individual user requests, as depicted in Fig.~\ref{fig:eval2}, which presents the empirical cumulative distribution function (CDF) for 10K requests across three environmental carbon intensities: 200, 300, and 400 gCO$_2$/kWh. The x-axis scales the CO$_2$ emissions of each request relative to executions on the \base{} system. Since we only show CO$_2$ per request, \cooopt{} is the best among all the schemes -- 80\% of requests have used less than 30\% of the \base{} carbon emission. When carbon intensity increases, \sol{}'s CDF moves closer and closer to \cooopt{}, indicating that \sol{}'s optimizer is adapting to the regional carbon intensity since the gain from using more concise directives gets amplified by higher carbon intensities (Sec.~\ref{sec:bkgd}). Specifically, when carbon intensity is 200 gCO$_2$/kWh, 40\% of \sol{}'s requests have used less than 40\% of the carbon footprint than \base{}; when it increases to 400 gCO$_2$/kWh, about 75\% of \sol{}'s requests have less than 40\% of \sol{}'s carbon footprint. Unlike \cooopt{} and \static{}, which do not adjust based on carbon intensity and thus maintain constant CDF curves, \sol{} exhibits a dynamic adaptation, aligning it closely with the \oracle{} benchmark in a per-request analysis.

\begin{figure}[t]
    \centering
    \includegraphics[scale=0.45]{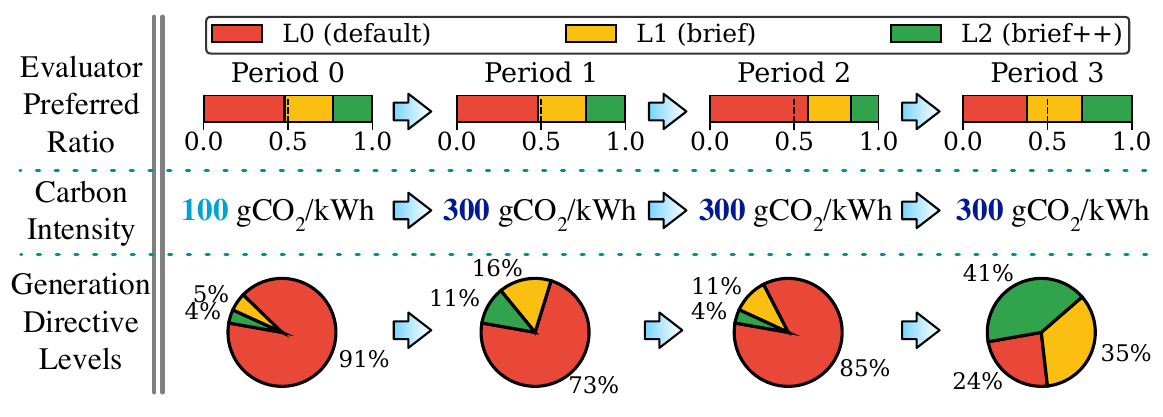}%\newline
    \vspace{1mm}
    \hrule
     \vspace{1mm}
    \caption{\sol{} configures the generation directive levels (represented as pie charts) according to quality evaluator preference and carbon intensity.}
    % \vspace{-0.3cm}
    \label{fig:eval3}
\end{figure}

Fig.~\ref{fig:eval3} illustrates \sol{}'s adaptive use of generation directive levels across different scenarios, represented through pie charts. During period 0, on average, the carbon intensity is 100 gCO$_2$/kWh and nearly half of the evaluations prefer the L0 directive. Consequently, \sol{} allocates L0 to 91\% of the prompts in total, reflecting its high preference rate. In period 1, with rising carbon intensity, there's a noticeable shift toward employing more L1 and L2 directives, aligning with environmental considerations. Period 2 presents unchanged carbon intensity but altered user preferences, leading to an increased preference for L0 by the evaluator and a corresponding adjustment in \sol{}'s directive assignments from 73\% to 85\% toward L0. Finally, in period 3, a significant change in user behavior emerges, showing a clear preference for L1 and L2 directives. This shift, coupled with the benefits of carbon savings at elevated carbon intensities, leads \sol{} to primarily assign L1 and L2 directives.

\begin{figure}[t]
    \centering
    \includegraphics[scale=0.45]{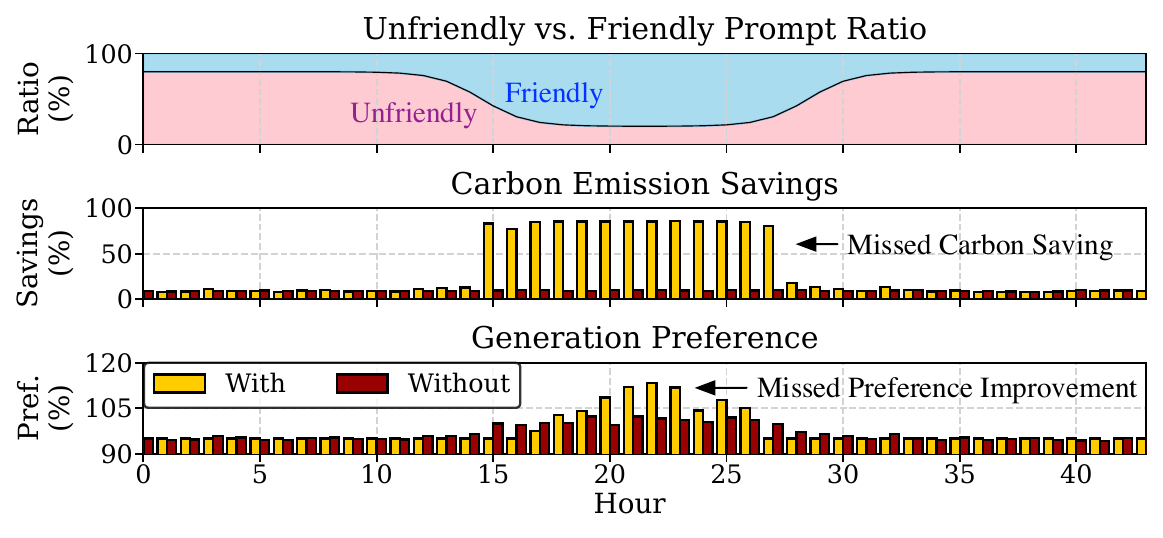}%\newline
    \vspace{1mm}
    \hrule
     \vspace{1mm}
    \caption{Without the offline evaluator, \sol{} misses the chance to leverage requests amenable to concise directive levels, thus forfeiting potential benefits in carbon savings and generation preference simultaneously.}
    % \vspace{-0.3cm}
    \label{fig:eval6}
\end{figure}

The offline quality evaluator is key to \sol{}'s effectiveness as we explain in Fig.~\ref{fig:eval6}. To show the necessity of the quality evaluator, we select \sol{}-friendly prompts which are prompts whose shorter responses are on average more preferred by the auto-evaluator than their default responses, and mix them with unfriendly prompts (shorter responses are less preferred by auto-evaluator than default responses). Over time, we vary the proportion of these two types of prompts, and we can observe that when the portion of friendly is high, \sol{} without the evaluator will miss out on the opportunity to save more carbon while achieving higher evaluator preference at the same time. As we can see around hour 22, the normalized preference is above 100\%, meaning the auto-evaluation LLM prefers \sol{}'s generation over the default generation more than 50\% of the time. 
The offline evaluator's low carbon overhead is also a key reason why \sol{} can save so much carbon with the evaluator, as we discuss next. 

\begin{figure}[t]
    \centering
    \includegraphics[scale=0.51]{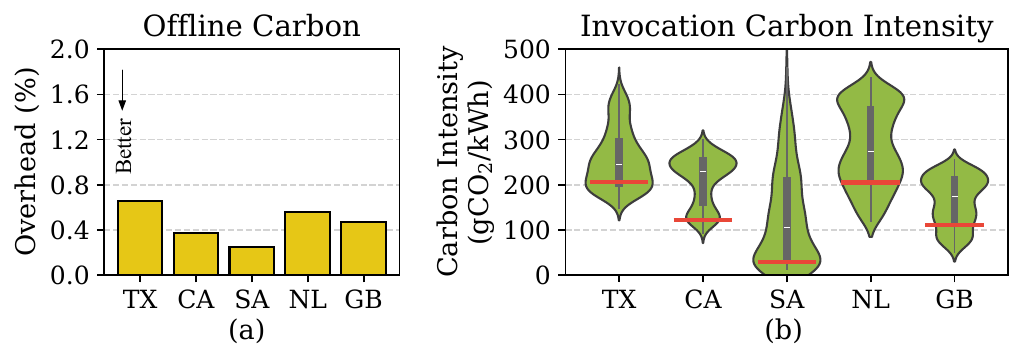}%\newline
    \vspace{1mm}
    \hrule
     \vspace{1mm}
    \caption{(a). \sol{}'s offline evaluator has negligible carbon emission overhead. (b). Violin plot of regional carbon intensity distribution, and the carbon intensity where \sol{} invokes offline evaluation (marked as red line).}
    % \vspace{-0.3cm}
    \label{fig:eval5}
\end{figure}

In Fig.~\ref{fig:eval5} (a), we show the carbon overhead of \sol{}'s offline evaluator. Since GPT-4 is only accessible from third-party API, we use the following numbers to estimate the offline evaluation carbon footprint. GPT-4 is speculated to use a mixture-of-experts (MoE) architecture, and during inference, only one expert is active. Thus, the model size is equivalent to one expert that has 220B parameters, which can be hosted on 16 A100 GPUs. With the measured average API accessing time of 500ms, we assume all 16 GPUs are running at max power (250W), under no network delay and no batched processing. Despite our conservative estimation where in reality the GPU generation time is much shorter than 500ms (network latency, pre- and post-processing) and multiple requests can be processed simultaneously in a batch, the overhead in Fig.~\ref{fig:eval5}(a) serving 30 requests per second (RPS)~\cite{kwon2023efficient} is still well below 1\% for all regions. The negligible carbon impact stems from (i) strategically timing evaluations to coincide with periods of low carbon intensity as shown in Fig.~\ref{fig:eval5} (b), and (ii) configuring the request to the auto-evaluation LLM in such a way that it generates only a minimal number of tokens for assessment, as detailed in Sec.~\ref{sec:implement}.

\subsection{Robustness and Implications}

\begin{figure}[t]
    \centering
    \includegraphics[scale=0.51]{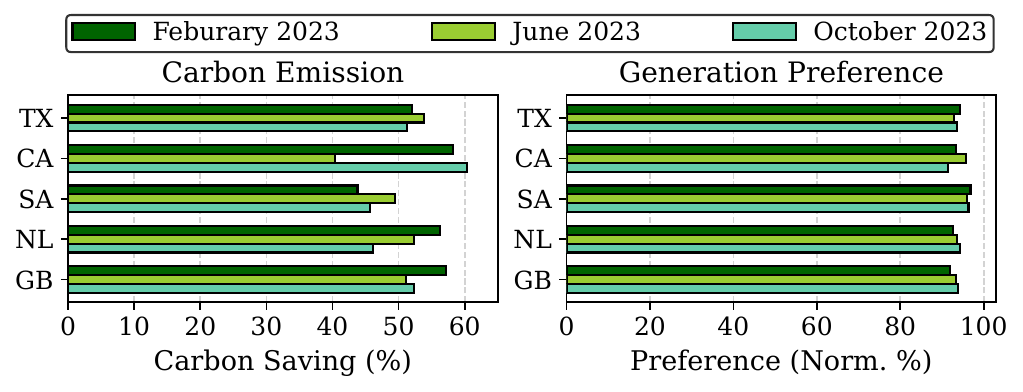}%\newline
    \vspace{1mm}
    \hrule
     \vspace{1mm}
    \caption{\sol{} remains effective in all geographical regions during different seasons.}
    % \vspace{-0.3cm}
    \label{fig:eval7}
\end{figure}

Finally, we assess the robustness of \sol{} and its broader implications. Fig.~\ref{fig:eval7} presents an evaluation of \sol{} across various periods of 2023, demonstrating its consistent efficacy across different seasons. \sol{} consistently enables the inference server to achieve over 40\% carbon emission savings while sustaining high levels of generation quality.

\begin{figure}[t]
    \centering
    \includegraphics[scale=0.405]{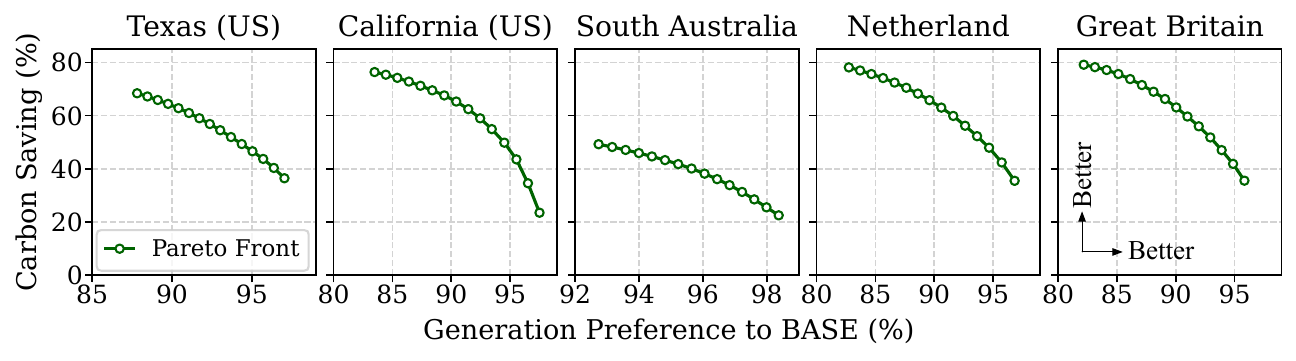}%\newline
    \vspace{1mm}
    \hrule
     \vspace{1mm}
    \caption{Pareto front of \sol{} across geographical regions when varying the preference coefficient. \sol{} can still reduce 40\% of carbon emission under strict generation preference constraints.}
    % \vspace{-0.3cm}
    \label{fig:eval8}
\end{figure}

\sol{} offers operators the ability to balance carbon savings against quality through the adjustable parameter $\xi$. Fig.~\ref{fig:eval8} illustrates the Pareto front demonstrating the trade-off between carbon savings and normalized generation preference as $\xi$ is varied. Remarkably, even when tightening the generation preference criterion to 95\% (indicating the evaluator prefers \sol{}'s generation over the default 48.7\% of the time), \sol{} consistently secures over 40\% carbon savings across all regions.

\sol{} stands as the inaugural approach to utilizing generation directives for configuring generative LLM inference, with a particular emphasis on advancing sustainability within GenAI by addressing carbon emissions. This strategy opens up extensive possibilities beyond its current focus. For instance, using generation directives can significantly enhance LLM inference throughput, thereby reducing the number of GPU servers needed to achieve specific rates of requests per second (RPS). This efficiency translates into reduced capital expenses for building LLM inference infrastructure and lowers the embodied carbon associated with manufacturing the GPU servers.
\section{Related Work}
\label{sec:related}

\noindent\textbf{Sustainable computing. } With the recent rise of interest and urgency toward reducing the carbon footprint of information technology, Totally Green~\cite{chang2012totally} and ACT~\cite{gupta2022act} introduced carbon modeling frameworks from system and architecture perspective. Based on the carbon modeling, Sustainable AI~\cite{wu2022sustainable}, Sustainable HPC~\cite{li2023toward}, and Chien et al.~\cite{dietrich2022navigating} have explored various carbon trade-offs in designing and operating large-scale computer systems. Various works have analyzed the recent trend and the future of AI development's impact on carbon emission~\cite{patterson2021carbon,patterson2022carbon,schwartz2020green,strubell2019energy,anderson2023treehouse}. \sol{} is motivated by these works and takes the effort a step further to LLM inference application. While systems like Ecovisor~\cite{souza2023ecovisor}, Dodge et al.\cite{dodge2022measuring}, Clover\cite{li2023clover}, and Carbon Explorer~\cite{acun2023carbon} have been designed to adapt to varying carbon intensities, they have not been specifically optimized for LLM inference workloads. Luccioni et al~\cite{luccioni2023estimating} and Chien et al.~\cite{chien2023reducing} have characterized the carbon profile and challenges from LLM inference, while \sol{} has designed an end-to-end framework that leverages generation directives to address these climate challenges. 
\vspace{2mm}

\noindent\textbf{Large language model inference. } Generative LLM inference is distinct from conventional ML inference due to its substantial parameter size, the extensive use of self-attention operations, and the autoregressive generation process. As such, the optimization strategies commonly applied in general ML inference contexts~\cite{romero2021infaas,li2023clover,wan2020alert,wang2023tabi,choi2022serving,li2023kairos,crankshaw2017clipper} are not ideally suited for generative LLMs. This gap has spurred the creation of dedicated LLM inference serving frameworks~\cite{cui2023optimizing,aminabadi2022deepspeed,zhou2022pets,miao2023spotserve,sheng2023fairness}, with notable examples like Orca~\cite{yu2022orca}, which introduces iteration-level batching, and vLLM~\cite{kwon2023efficient}, known for its efficient KV cache management via paging. \sol{} is engineered for compatibility with these specialized frameworks, with its effectiveness stemming not from the particulars of the inference server's setup but from the strategic use of token generation directives.

The surge in LLM inference popularity has prompted a diverse range of research on performance and memory optimization, exploring strategies like sparsity and pruning~\cite{liu2023deja,frantar2023sparsegpt}, speculative decoding~\cite{leviathan2023fast,chen2023accelerating}, GPU kernel tiling and fusion~\cite{dao2023flashattention,zheng2023pit}, disk offloading~\cite{sheng2023flexgen,rajbhandari2021zero}, and mixture-of-experts approaches~\cite{li2023accelerating,xue2024moe}. These advancements are crucial for facilitating the deployment of larger LLMs to a broader audience. However, the environmental implications of these technologies are equally important. While LLMCarbon~\cite{faiz2023llmcarbon} offers carbon footprint predictions to help researchers gauge the environmental impact of LLMs prior to training, \sol{} stands out as the first work to tackle the carbon footprint challenge of generative LLM inference using a novel generation directive mechanism.
\section{Conclusion}
\label{sec:conclude}

This paper introduced \sol{}, an innovative framework designed to enhance the sustainability of generative AI by creating a carbon-aware inference service for generative language models. Utilizing the novel concept of generation directives, \sol{} significantly optimizes the carbon footprint associated with LLM inference. Our evaluation, conducted using a Llama2 inference server and a GPT-4 quality evaluator, demonstrates that \sol{} can reduce the carbon footprint of inference activities by over 40\% in various global regions. Through the development of \sol{}, we seek to pave the way for a greener future in generative AI, stimulating further research into minimizing the environmental impacts in the era of rapid AI advancements.

\section*{Acknowledgments}
This material is based upon work supported by the Assistant Secretary of Defense for Research and Engineering under Air Force Contract No. FA8702-15-D-0001, and United States Air Force Research Laboratory Cooperative Agreement Number FA8750-19-2-1000. Any opinions, findings, conclusions, or recommendations expressed in this material are those of the author(s) and do not necessarily reflect the views of the Assistant Secretary of Defense for Research and Engineering, or the United States Air Force. The U.S. Government is authorized to reproduce and distribute reprints for Government purposes notwithstanding any copyright notation herein.

\bibliographystyle{IEEEtran}
\bibliography{refs}

\end{document}